\documentclass[a4paper,12pt]{article}
\usepackage{graphicx} %use graph format
\usepackage{epstopdf}

\begin{document}

%%-move to normal A4-%%
\hoffset = -1truecm \voffset = -2truecm \baselineskip = 10 mm

\title{\bf A simplified connection between constituent quark and parton}

\author{Wei Zhu$^1$ and Fan Wang$^2$
\\
\normalsize $^1$Department of Physics, East China Normal University,
Shanghai 200241, China \\
\normalsize $^2$Department of Physics, Nanjing University,
Nanjing,210093, China\\
}

\date{}

\newpage

\maketitle

\vskip 3truecm

\begin{abstract}

    We propose a simple way to connect Gell-Mann constituent quark model
and Feynman parton model for the nucleon. Thus, we can dynamically
understand a large amount of data for high energy hadronic processes
starting from the first QCD principle.

\end{abstract}

{\bf keywords}: Constituent quark model, parton model, perturbative
QCD, nonperturbative QCD

\vskip 1truecm

\newpage

    The consistent quark model (CQM) and the parton model (PM) successively described
two different faces of a true nucleon: one is that the nucleon is
consisted of limited number of massive (dressed) quarks and other
regards the proton as a cluster of an infinite number of massless
partons (quarks and gluons). Unfortunately, their connection is
still an open question.

    Experiments at high energy have accumulated a large amount of data about
parton distributions in the nucleon. According to the factorization
theorem [1], if we have measured parton distributions over a range
$x_0\leq x\leq 1$ at one value of $Q^2_0$, we can use perturbative
QCD (pQCD) to predict parton distributions for $Q^2>Q^2_0$ over the
same range in the Bjorken variable $x$. The variation of the parton
distributions with $Q^2$ is known as their evolution. A selected
value of $Q^2_0$ in this example is called as the factorization
scale and the parton distribution $f(x,Q^2_0)$ as the starting
parton distribution. Usually, $Q^2_0>1~GeV^2$ and $f(x,Q^2_0)$
contains limited number of intrinsic (valence-like) quarks and
infinite gluons and seq quarks. The application of the factorization
theorem further assumes there is a special factorization scale
$\mu^2$, where all the long-distance dependence resides in the
nonperturbative input parton distributions $f_{IP}(x,\mu^2)$, while
all short-distance ($Q^2$) dependence is in the evolvable
perturbative parton distributions. According to this definition, the
input parton distribution relates to a finite number of intrinsic
partons. We will show that only the input parton distribution
suitable to connect with the following constituent quark
distribution.

    Recently the nonperturbative QCD research (Schwinger-Dyson equation
[2] and light front QCD [3]) reproduced the Gell-Mann picture about
hadrons, which are mainly composed of three or two massive
constituent quarks, or adding a limited number of intrinsic
quark-antiquark pair. The results can describe the hadron spectra
and the various form factors.

    A following key step is to connect the wave function of constituent quarks with the input parton
distribution. Once we found a "bridge" connecting these two
distributions, one can predict a lot of hadronic processes at high
energy starting from the first principle. Although several of
starting distributions for partons at $Q^2_0>1~GeV^2$ have been
extracted from the experimental data with the linear DGLAP equation
at the twist-2 level [4-6], however, these starting distributions
always mix with the infinite number of gluons and sea quarks, they
cannot correspond to the finite intrinsic components of the input
distributions.

    A nucleon is
naively consisted of three constituents without the probing scale
$Q^2$. A natural attempt beginning from 1976, is to assume that the
nucleon has three valence quarks at a low starting point $\mu^2$
(but still in the perturbative region $\alpha_s(\mu^2)/2\pi <1$ and
$\mu>\Lambda_{QCD}$), and the gluons and sea quarks are
radioactively produced at $Q^2>\mu^2$ [7-9]. However, such natural
input fails due to overly steep behavior of the predicted parton
distributions at the small $x$. Instead of this natural input, Reya,
Gl\"{u}ck and Vogt (GRV) [10] added the valance-like sea quarks and
valence-like gluon distributions to the input parton distribution.
These valence-like components can slow down the evolution of the
DGLAP equation at low $Q^2$ and reach agreement with experimental
results, since the evolution region of the valence-like
distributions is sizeably larger. However, either proton or neutron
is not a hybrid hadron containing the intrinsic gluons. Therefore,
it is difficult to connect the GRV-input with the constituent quark
distribution.

      As we know that the contributions of the parton recombination
corrections become important at $Q^2<1~ $GeV$^2$, which are
neglected in the DGLAP equation. The negative corrections of the
parton recombination should slow down the parton evolution. These
nonlinear effects can be calculated by pQCD at the twist-4 level.
Such nonlinear corrections of the gluon recombination to the DGLAP
equation were firstly derived by Gribov, Levin and Ryskin [11] and
Mueller and Qiu [12] in the double leading logarithmic
($DLL(1/x,Q^2$)) approximation. This evolution equation was
re-derived by Zhu, Ruan and Shen [13-15]. The ZRS-version of this
equation restores the momentum conservation, takes the leading
logarithmic $(LL(Q^2))$ approximation and includes all parton
recombination, therefore, it can naturally connect with the DGLAP
equation and works in the whole $x$ range. Based on the ZRS version
of this equation, the possible available input distributions for the
nucleon have been proposed [16-19], where the shadowing effect
replaces the corrections of the valence-like gluon, the input
contains only three valence quarks if without the flavor-asymmetric
sea components, or adds the small amount of valence-like sea quarks
if considering the flavor-asymmetry in the sea quark distributions.

    However, both the GRV-input and the input based on the GLR-MQ-ZRS equation encountered the following disputes.
(i) Is pQCD valid at $Q^2<1~GeV$ ? (ii) What is the conversion
between the constituent quark and the parton at low $Q^2$? The
purpose of this letter is try to answer the above questions. We
first focus on the corrections of the intrinsic quark mass to the
QCD evolution equation when the equation works at the low $Q^2$
range. We find that the mass-effect of the intrinsic partons may
freeze the pQCD evolution at $Q^2<M^2_{eff}$, $M_{eff}\sim 300~MeV$
is a simplified common mass-scale for the dressed light-quarks and
dressed gluons. While this mass-effect gradually disappears at
$Q^2>1~GeV$ even if $M_{eff}\neq 0$. The parton distributions in the
transition range $M^2_{eff}<Q^2<1~GeV^2$ will evolve along a special
path $\tilde{Q}^2=Q^2+M^2_{eff}$.

    A main difference between the constituent quark and the parton is that the former is massive $M\sim m_p/3$, while the later
is regarded as massless. The mass $M$ may be arisen from the
propagation of  bare-quark (even bare-gluon) in the strongly-coupled
non-Abelian gauge field [20]. The QCD evolution equation is derived
in the perturbative domain, where all partons are massless. We
consider the corrections of the mass effect to the GLR-MQ-ZRS
equation at the low $Q^2$ range. For the sake of simplicity, we only
write it for the flavor singlet quarks, their evolution reads

$$Q^2\frac{dxq^s(x,Q^2)}{dQ^2}$$
$$=\frac{\alpha_s(Q^2)}{2\pi}[P_{qq}\otimes q^s+P_{qg}\otimes g]$$
$$-\frac{\alpha_s^2(Q^2)}{4\pi R^2 Q^2}\int_x^{1/2}\frac{dy}{y}x
P_{gg\rightarrow q}(x,y)[ yg(y,Q^2)]^2$$
$$+\frac{\alpha_s^2(Q^2)}{4\pi R^2 Q^2}\int_{x/2}^x\frac{dy}{y}x
P_{gg\rightarrow q}(x,y)[ yg(y,Q^2)]^2, (if~x\le 1/2),$$

$$Q^2\frac{dxq^s(x,Q^2)}{dQ^2}$$
$$=\frac{\alpha_s(Q^2)}{2\pi}[P_{qq}\otimes q^s+P_{qg}\otimes g]$$
$$+\frac{\alpha_s^2(Q^2)}{4\pi R^2 Q^2}\int_{x/2}^{1/2}\frac{dy}{y}x
P_{gg\rightarrow q}(x,y)[ yg(y,Q^2)]^2,(if~1/2\le x\le 1).
\eqno(1)$$The DGLAP splitting function for $q(l)\rightarrow
q(k)+g(l')$ is [5]

$$P_{qq}\frac{dk^2_T}{k^2_T}=\frac{E_k}{E_l}\vert M_{l\rightarrow kl'}\vert^2
\left[\frac{1}{E_l-E_k-E_{l'}}\right]^2\left[\frac{1}{2E_k}\right]^2
\frac{d^3l'}{(2\pi)^32E_{l'}}.\eqno(2)$$The momentum of partons are
written in the infinite momentum frame, they are

$$l=[x_1p,0,x_1p];~k=\left[x_2p+\frac{k^2_T}{2x_2p},k_T, x_2p\right];~l'=\left[x_3p+\frac{k^2_T}{2x_3p},-k_T,
x_3p\right]. \eqno(3)$$

    The modification of the massive propagator leads to the following change in the $k_T$ dependent part of $P_{qq}$,

$$\frac{k^2_Tdk^2_T}{k^4_T}\rightarrow \frac{k^2_Tdk^2_T}{(k^2_T+M_{eff}^2)^2}\approx \frac{d(k^2_T+M^2_{eff})}{k^2_T+M_{eff}^2}
\equiv\frac{d\tilde{k}^2_T}{\tilde{k}^2_T}. \eqno(4)$$The
$k_T^4$-factor in the denominator of first formula origins from the
energy defect (i.e., the propagator in the time ordered perturbative
theory form) in Eq. (2), therefore, it is replaced by
$k^2_T+M^2_{eff}$. While the $k_T^2$-factor in the numerator of
first formula is the result of the matrix $M_{l\rightarrow kl'}$,
which is consisted of bare-vertex and is irrelevant to $M_{eff}$.
The approximation in Eq. (4) may cause about maximum double
deviation from a correct increment of the evolution at $k^2_T\simeq
M^2_{eff}$, although this deviation fast disappears at
$k^2_T\gg\mu^2$. However, the evolution increment at such lower
$k_T$-scale is small since it is thin gluon environment. Therefore,
we neglect the above deviation.

    A similar result satisfies the recombination function in Eq. (1). Thus we have [12]

$$P_{gg\rightarrow q}\frac{dk^2_T}{k^4_T}=\frac{E_k}{E_l+E_2}\vert M_{p_1p_1\rightarrow kl'}\vert^2
\left[\frac{1}{E_1+E_1-E_k-E_{l'}}\right]^2\left[\frac{1}{2E_k}\right]^2
\frac{d^3l'}{(2\pi)^32E_{l'}}.\eqno(5)$$Corresponding to Eq. (4),

$$\frac{(k^2_T/k^2_T)dk^2_T}{k^4_T}\rightarrow \frac{k^2_Tdk^2_T}{(k^2_T+M_{eff}^2)^3}\approx \frac{d(k^2_T+M^2_{eff})}{(k^2_T+M_{eff}^2)^2}
\equiv\frac{d\tilde{k}^2_T}{\tilde{k}^4_T}. \eqno(6)$$Note that
$k^2_T/k^2_T\rightarrow k^2_T/(k^2_T+M^2_{eff})$ since the matrix
$\vert M_{p_1p_1\rightarrow kl'}\vert^2$ contributes a pair of
massive propagators. According to work [21], the the modified
running coupling due to the mass effect is
$\alpha_s(k^2_T+M^2_{eff})$. The suppression of the running-coupling
near the $\mu^2$-scale helps improve the convergence of perturbative
expansion.

        The renormalization group theory is the basis of the standard  QCD evolution equation. Comparing
with the renormalization group equation for the moments of the
structure function, one needs to set $k^2_T\rightarrow Q^2$ in the
splitting function (2) and recombination function (5) when
constructing Eq. (1) [6]. Only in this way, Eqs. (2) and (5) can
play the role of the evolution kernels. Thus, the GLR-MQ-ZRS
equation (1) including massive corrections Eqs. (4) and (6) can be
rewritten as

$$\overline{Q}^2\frac{dxq(x,\overline{Q}^2)}{d\overline{Q}^2}$$
$$=\frac{\alpha_s(\overline{Q}^2)}{2\pi}[P_{qq}\otimes q+P_{qg}\otimes g]$$
$$-\frac{\alpha_s^2(\overline{Q}^2)}{4\pi R^2 \overline{Q}^2}\int_x^{1/2}\frac{dy}{y}x
P_{gg\rightarrow q}(x,y)[ yg(y,\overline{Q}^2)]^2$$
$$+\frac{\alpha_s^2(\overline{Q}^2)}{4\pi R^2 \overline{Q}^2}\int_{x/2}^x\frac{dy}{y}x
P_{gg\rightarrow q}(x,y)[ yg(y,\overline{Q}^2)]^2, (if~x\le 1/2),$$

$$\overline{Q}^2\frac{dxq(x,\overline{Q}^2)}{d\overline{Q}^2}$$
$$=\frac{\alpha_s(\overline{Q}^2)}{2\pi}[P_{qq}\otimes q+P_{qg}\otimes g]$$
$$+\frac{\alpha_s^2(\overline{Q}^2)}{4\pi R^2 \overline{Q}^2}\int_{x/2}^{1/2}\frac{dy}{y}x
P_{gg\rightarrow q}(x,y)[ yg(y,\overline{Q}^2)]^2,(if~1/2\le x\le 1).
\eqno(7)$$which evolves with $Q^2$ along a following special path

$$\overline{Q}^2=Q^2+M^2_{eff}, \eqno(8)$$where $M_{eff}\simeq 300~MeV$ is a common scale
for either the constituent quarks and dressed gluons. Figure 1
presents the relation between $\overline{Q}^2\sim Q^2$. On can find
that $\overline{Q}^2$ gradually approaches to a lower limit
$M^2_{eff}$ with $Q^2\rightarrow 0$. We freeze the perturbative QCD
evolution at $Q^2<\mu^2$ because where $d\ln \overline{Q}^2\simeq
0$. This is consistent with the definition of the factorization
scale $\mu^2$.

    A complete structure function for the nucleon can be perturbatively expanded on the
twist

$$F_N(x,Q^2)=F_N^{(2)}(x,Q^2)+\sum_{n=2}^{\infty}F^{(2n)}_N(x,Q^2),\eqno(9)$$where
the leading twist $F^{(2)}_N$ is generally described by the parton
model. As we known that in history the parton description was once
considered valid only at the Bjorken limit $Q^2\gg m^2_N$, for say,
at last at $Q^2>10~GeV^2$. However, the experimental data show that
the DGLAP equation at the twist-2 level is already effective enough
at $Q^2> 1~GeV$. This fact is called as the precocity of the parton
model due to asymptotic freedom of QCD, i.e., $\alpha_s(Q^2)/\pi\ll
1$ at $Q^2>1~GeV^2$. Taking this result, a following twist-4
correction is either necessary and sufficient in the range
$0.1~GeV^2<Q^2<1~GeV^2$. Besides, the suppression of the
running-coupling with the mass-effect $\alpha_s(Q^2+M^2_{eff})$ at
low $Q^2$ helps improve the convergence of perturbative expansion.
If we further consider that the evolution is freezed at
$Q^2<M^2_{eff}$, where the complex nonperterbative effects are
covered, the questions about the validity of the GLR-MQ-ZRS equation
at the range $Q^2>\mu^2$ have a positive answer (see Fig. 2).

     In order to check above our understanding, we take an input parton distribution based on the GLR-MQ-ZRS
equation from Ref. [16]

$$xu_{IP}(x,\mu^2)=\frac{2}{B(1.98,3.06)}x^{1.98}(1-x)^{2.06},$$
$$xd_{IP}(x,\mu^2)=\frac{1}{B(1.31,5.8)}x^{1.31}(1-x)^{4.8},  \eqno(10)$$which is extracted from a globe fitting
by the GLR-MQ-ZRS equation; $B$ is Beta function.  Note that in work [17] the suppression of
$\alpha_s$ at $\mu^2<Q^2<1~GeV^2$ is neglected but the evolution is freezed at
$Q^2<\mu^2=0.064~GeV^2$ (see the broken curve in Fig. 1). For the sake of simplicity,
we assume that $\mu^2=M^2_{eff}$ since they are of the same order of magnitude.

    Recently, the BLFQ collaboration [22,23] gives the constituent quark distribution $f_{CQ}(x)$
in the proton using the light-front (LF) QCD. We take it as an
example of the constituent quark model and plot two distributions
$f_{CQ}(x)$ and $f_{ZRS}(x,\mu^2)$ ($f=u,d$) in Fig. 3.  One can
find that two distributions are close, however, there is a not
negligible difference between them. It seems a part of momentum
fraction transfers from $d$-quark to $u$-quark in the proton at
$Q^2<\mu^2$. This process is nonperturbative since the perturbative
evolution has been freezed below the factorization scale $\mu$. We
try to understand it as follows. The asymmetry $u$- and $d$-Coulomb
potential $V_{ud}+V_{uu}>2V_{du}$ in the proton of the constituent
quark model is negligible comparing with the strong QCD interaction
since $\alpha_{em}\ll\alpha_s$. However, a quark is knocked out a
nucleon by impulse at $Q^2=\mu^2$ in deep inelastic scattering, it
transits from bound state to a free one. Although we don't know the
details of the proton fragmentation, according to the parton model,
the struck quark keeps its original distribution if without extra
interaction. However, the interaction of the Coulomb field in this
case is highlighted due to the QCD de-confinement. The average
momentum of the proton will redistributed between $u$- and $d$-CQs,
i.e., a part of momentum fraction will transfer from $d$-quark to
$u$-quark since the total momentum of quarks are conservation. We
write it as

$$2<u(\mu^2)>_2=2<u>_2+\Delta x, $$

$$<d(\mu^2)>_2=<d>_2-\Delta x,\eqno(11) $$where $<...>_2$ is the second moment of the distribution.
This nonperturbative deformation of the quark distributions keeps
the number of the quarks and their total momentum,

$$\int_0^1dxx\sum_{f=u,d} f_{CQ}(x)=\int_0^1dxx\sum_{f=u,d} f_{IP}(x,\mu^2)=1,\eqno(12)$$

$$\int_0^1dxu_{CQ}(x)=\int_0^1dxu_{IP}(x,\mu^2)=2,\eqno(13)$$and

$$\int_0^1dxd_{CQ}(x)=\int_0^1dxd_{IP}(x,\mu^2)=1.\eqno(14)$$
The simple mathematical form of Eq.(10) allows us to determine the
deformation form using the minimum free parameters. The Regge
exchange dominates the nonperturbative input distribution at
$x\rightarrow 0$ [24]. The exchanged Regge trajectory is determined
by the target quantum numbers, which are irrelevant to the
deformation of the distributions. Therefore, we assume that two
power indexes 1.98 and 1.31 in Eq.(10) are almost invariant in the
deformation. Thus, we have  only a free parameter $\Delta x$. We
take $\Delta x=0.08$ and get the constituent quark distribution
before the deformation $f_{CQa}(x)$ using Eq.(10). The results are
presented in Fig. 3 (dashed curves), they are parameterized as

$$xu_{CQa}(x)=\frac{2}{B(1.98,3.63)}x^{1.98}(1-x)^{2.63},$$
$$xd_{CQa}(x)=\frac{1}{B(1.31,3.14)}x^{1.31}(1-x)^{2.14}.  \eqno(15)$$ One can find that

$$u_{CQa}(x)\simeq u_{CQ}(x);~~d_{CQa}(x)\simeq d_{CQ}(x).\eqno(16)$$

    To test the validity of the above treatment about the deformation of the quark distributions,
we take a similar example in Fig. 4, where the distributions
$f_{NJL}(x)$ is provided by the Nambu-Jona-Lasinio model [25] and
$\Delta x=0.06$. The corresponding constituent quark distributions
in the proton before the deformation are

$$xu_{CQb}(x)=\frac{2}{B(1.98,3.48)}x^{1.98}(1-x)^{2.48},$$
$$xd_{CQb}(x)=\frac{1}{B(1.31,3.47)}x^{1.31}(1-x)^{2.47},  \eqno(17)$$ They are consistent
with the constituent quark distributions
in the $NJL$ model, i.e.,

$$u_{CQb}(x)\simeq u_{NJL}(x);~~d_{CQb}(x)\simeq d_{NJL}(x).\eqno(18)$$The above discussions
also satisfy the quark distributions in meson. In this case $\Delta
x=0$ because the symmetry of the Coulomb energy
$V_{u\overline{d}}=V_{\overline{d}u}$ or
$V_{u\overline{s}}=V_{\overline{s}u}$, $s$ is strange quark.

    A more accurate fitting between two distributions needs to consider the improvement of the
QCD dynamic model and the higher-order corrections of perturbative
calculation. Besides, the above discussions take the three quark
approximation. We need consider the contributions of multi-quarks
state in the constituent quark model and the asymmetry intrinsic sea
quarks in the parton distributions. The GLR-MQ-ZRS equation has
prepared such input [15,16]. By the way, the measured structure
functions in deep inelastic scattering are not exactly the
contributions of the parton distributions at very low $Q^2$, which
is beyond the impulse approximation, the contributions of the
handbag-diagram representation of the virtual-photon-pion forward
Compton scattering amplitude should be noted when comparing the
results with the experimental data. For a detailed discussion, see
Ref. [26]. Thus, we realise a connection between a nonperturbative
quark model and the perturbative parton model for the nucleon.

    In summary, the constituent quark distribution and a set of input parton distribution in the proton are compared.
We find the nonperturbative deformation of the quark distribution
below the factorization scale $\mu^2$. A possible deformation
mechanism and the quark mass effect at the transfer range are
discussed. Based on the above discussions, a simplified connection
between the constituent quark model and the parton model is
established at the three quark configuration. The result is useful
to realize the connection of the constituent quark model and the
parton model.

\noindent {\bf Acknowledgments} We thank J.S. Lan, P. Liu, J.H.
Ruan, C.D. Roberts and X.P. Zhao for useful discussions and thank
S.Q. Xu for providing $f_{CQ}(x)$ in Fig.3. This work is supported
by the National Natural Science of China (No.11851303).

\newpage
\begin{figure}
    \begin{center}

        \includegraphics[width=0.8\textwidth]{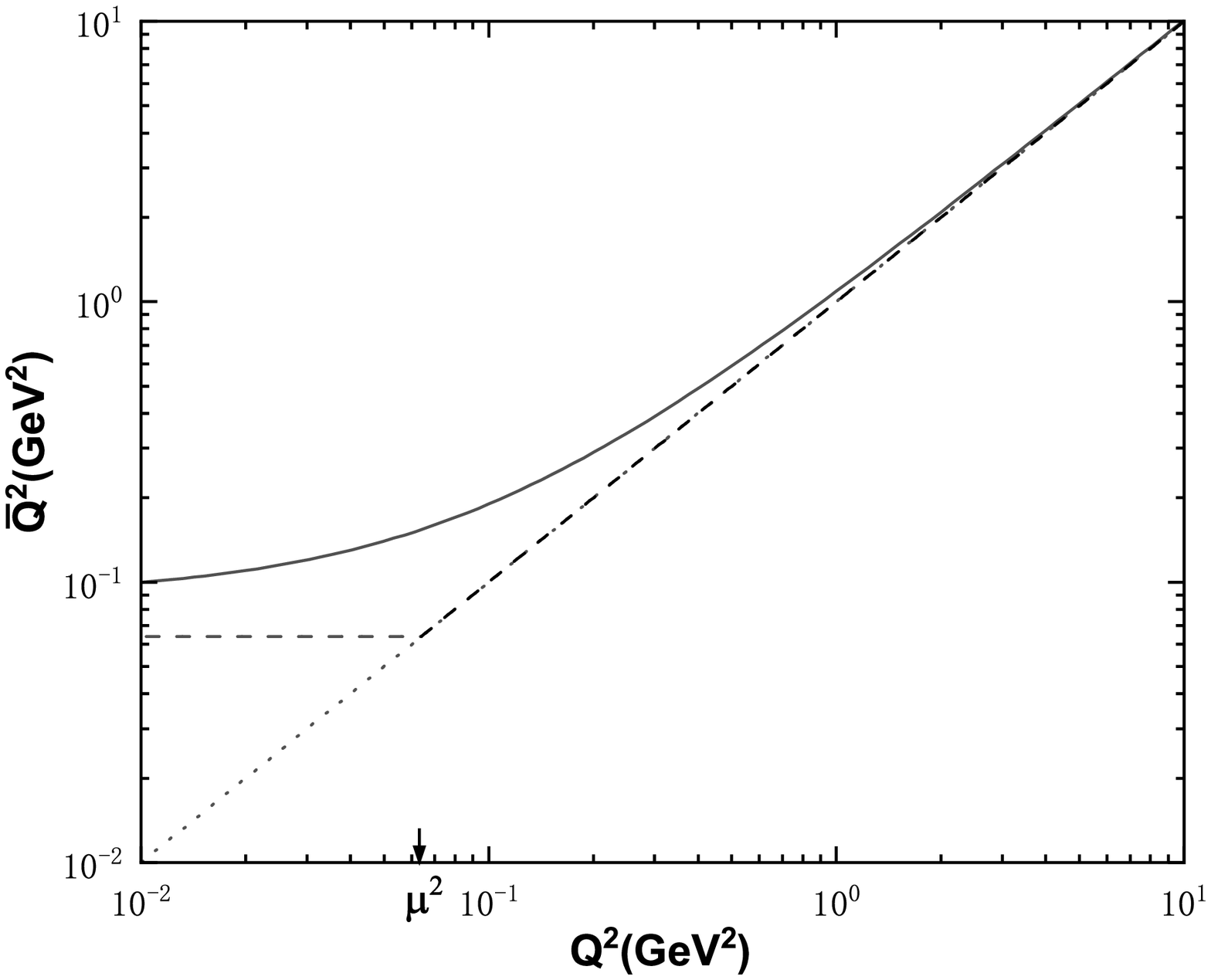}
        \caption{The new evolution path $\overline{Q}^2\sim Q^2$ due to the mass-effect.
        The result shows that the evolution is approximately saturated at $Q^2<\mu^2$. }\label{fig:1}

    \end{center}
\end{figure}

\newpage
\begin{figure}
    \begin{center}

        \includegraphics[width=0.8\textwidth]{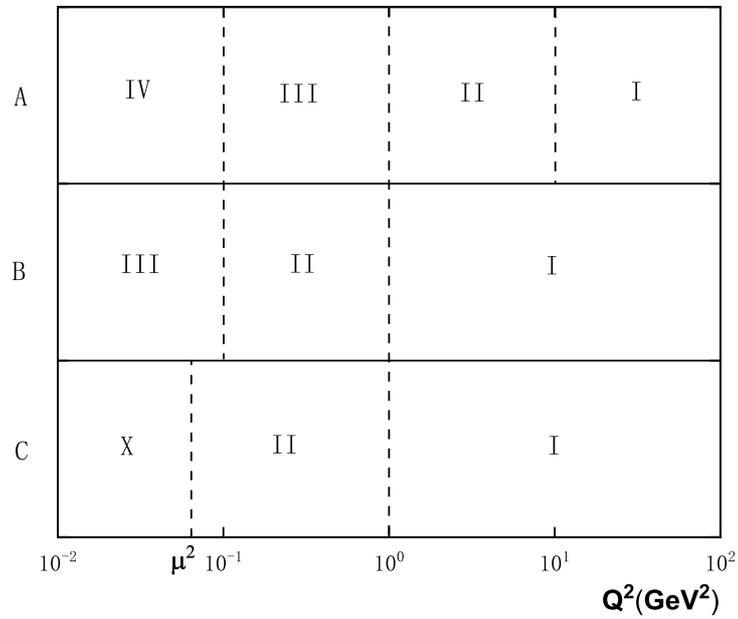}
        \caption{The schematic diagram for the applicable range of the perturbative expansion (9). (I)
    for $F^{(2)}(x)$, (II) for $F^{(2)}(x)+F^{(4)}(x)$, (III) $F^{(2)}(x)+F^{(4)}(x)+F^{(6)}(x)$, (X) where all perturbative evolutions are freezed. A: A naive
    expectation, B: Precocious parton model, C: This work.
         }\label{fig:2}

    \end{center}
\end{figure}

\begin{figure}
    \begin{center}

        \includegraphics[width=0.8\textwidth]{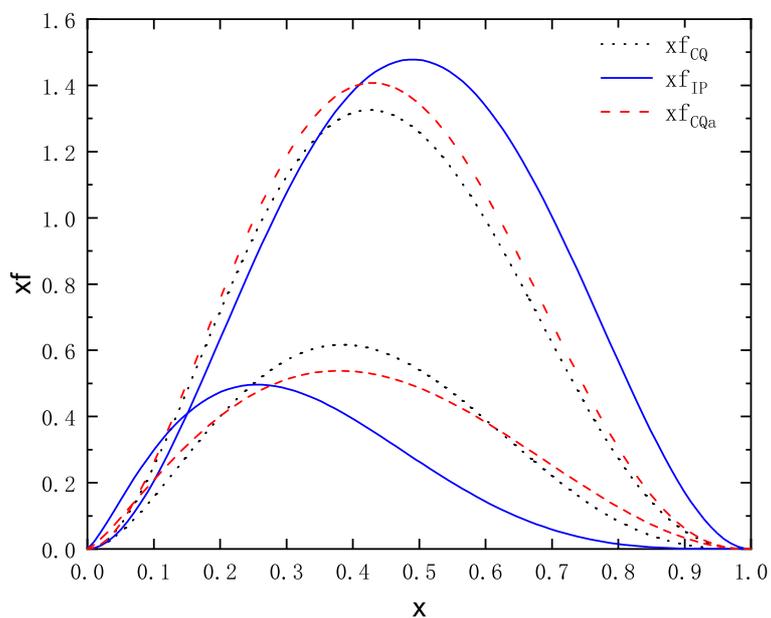}
\caption{The input parton distribution in the proton
$xf_{IP}(x,\mu^2)$ (solid curves) taken from Ref.[16]; Predicted
constituent quark distribution $xf_{CQa}(x)$ (dashed curves), which
compares with the constituent quark distribution $xf_{CQ}(x)$ (point
curves) in the LFQCD model [24]. A free parameter $\Delta x=0.08$.
The above set is the u-quark distribution and the following set is
the d-quark distribution.} \label{fig:3}

    \end{center}
\end{figure}

\begin{figure}
    \begin{center}

        \includegraphics[width=0.8\textwidth]{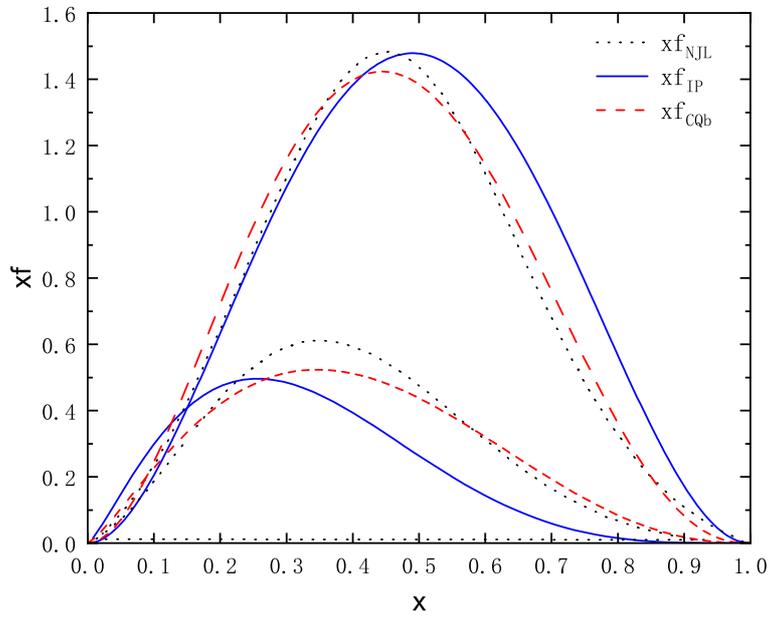}
\caption{Similar to Fig.3 but for the comparison with the
constituent quark distribution $xf_{NJL}(x)$ (point curves) in the
NJL model [26]. A free parameter $\Delta x=0.06$.} \label{fig:4}

    \end{center}
\end{figure}

\end{document}